# Durability of solid oxide electrolysis stack under dynamic load cycling for syngas production

Megha Rao, Xiufu Sun* xisu@dtu.dk, Anke Hagen

Department of Energy Conversion and Storage, Technical University of Denmark, Roskilde, 4000, Denmark

*Corresponding author.

## Abstract

A 6-cell solid oxide electrolysis stack was tested under $H_2O + CO_2$ co-electrolysis conditions. The cells used in the stack consisted of a nickel-yttria stabilized zirconia (Ni-YSZ) fuel electrode, YSZ electrolyte and lanthanum strontium cobaltite-gadolinium doped ceria (LSC-GDC) composite oxygen electrode. The aim of this study was to investigate the stack durability when operated under dynamic load conditions simulating a wind energy powered SOEC stack for synthesis gas production. The degradation of the stack was observed to be less than 1%/1000 h in terms of area specific resistance during the 1000 h operation. Detailed electrochemical analysis revealed a constant ohmic resistance, indicating intact contact in the stack. Only minor degradation was observed, mainly due to the fuel electrode process. The overall stack voltage degradation rate was 0.8%/1000 h.

## 1 Introduction

Europe has a strong commitment to reduce the emission of greenhouse gasses [1]. In order to achieve this goal, electricity production from renewable sources is strongly extended. Most matured technologies for renewable energy production are solar and wind. However, given the intermittent nature of power generation, balancing and storage technologies are needed. Pumped hydro, compressed air, and batteries are well-known energy storage technologies so far [2,3]. Furthermore, electrolysis is a promising solution for highly efficient large-scale energy storage, in particular high temperature electrolysis using solid oxide electrolysis cells (SOECs). Power-to-Gas (PtG)/Power-to-liquid (PtL) scenarios can be realized using SOECs, wherein $CO_2$ and steam can be combined in the electrolysis to form synthesis gas (syngas: $CO + H_2$). Syngas can be converted downstream to produce hydrocarbons using established catalytic processes [4–9]. Methane is of special interest owing to the availability of infrastructure for transport and storage in Europe [10].

One of the main challenges associated with the application of SOECs is the durability of the cells and stacks to reach lifetimes of 5–10 years. Durability tests have been carried out under constant current under steam and co-electrolysis conditions [11–15]. The cells have been tested at various operating parameters such as temperatures, gas compositions, current densities and voltages. For the State-of-the-art (SoA) Ni-YSZ fuel electrode supported SOECs, the main degradation has been attributed to the fuel electrode such as Ni coarsening, Ni migration and loss of percolation [16–20]. In addition, the oxygen electrode degrades owing to the activity of the oxygen evolution reaction in the electrode [21,22], which also depends on the structure and composition of the oxygen electrode. Moreover, impurities in the gas stream effect the cell performance and durability [14,23].

Stack tests with fuel electrode supported cells under co-electrolysis conditions were reported in the literature [24–27]. It was shown that the measured initial area specific resistance (ASR) was higher in co-electrolysis than steam electrolysis gas compositions. Furthermore, temperature plays an important role on both stack performance and degradation rate. The analysis of the co-electrolysis outlet gas composition at 750 °C revealed that the reverse water gas shift reaction plays an important role and that the outlet gas was close to the thermodynamic equilibrium composition [26,27]. Degradation mechanisms such as loss of Ni–Ni contact and Ni coarsening were reported after long term co-electrolysis operation under constant operating conditions [25].

Testing of SOECs are traditionally performed under galvanostatic conditions due to the ease of operation and data interpretation. A few tests have also been performed potentiostatically, at constant voltage, to understand the degradation of cells and the correlation between the testing modes and to approach realistic operating conditions at thermoneutral voltage [28,29]. Considering the intermittent energy supply/storage, it is of importance to operate the stack under dynamic load conditions and evaluate the degradation during such operation mode. In this work, we studied the operation of a 6-cell stack under dynamic load operation using a 24 h wind profile, which was repeated over a period of 1000 h. Detailed electrochemical characterization, with cell voltage evolution over time and electrochemical impedance spectra recorded prior to, under, and after dynamic durability test, was performed on individual cells in the stack, and the degradation of the cells under dynamic load cycle operation is presented.

## 2 Experimental

A stack with six cells produced by DTU Energy was assembled by SOLIDPower. The stack was pre-reduced when received from SOLIDPower. Cells with 30 μm thick composite LSC-CGO oxygen electrode, a 6–7 μm thick CGO barrier layer, a 10 μm thick YSZ electrolyte, a 12–16 μm thick Ni-YSZ fuel electrode and a 300 μm thick Ni-YSZ support layer were used in the stack (for cell details see Ref. [16]). The active area of one cell was 80 $cm^2$. The stack was tested in the setup shown in Fig. 1. The details are described elsewhere [30].

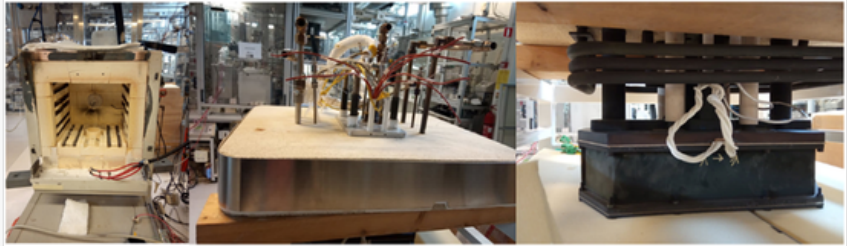

Fig. 1

Stack test setup with furnace and stack mounting images.

Initial characterization by electrochemical impedance spectroscopy (EIS) and iV curves (fingerprint) on each of the 6 cells of the stack was performed at 750 °C and 700 °C with the following fuel gas compositions supplied to the fuel electrode compartment:

1. 20% $H_2O$ + 80% $H_2$: 2.5 NL·min$^{-1}$ $H_2O$ and 10 NL·min$^{-1}$ $H_2$
2. 50% $H_2O$ + 50% $H_2$: 2.45 NL·min$^{-1}$ $H_2O$ and 2.45 NL·min$^{-1}$ $H_2$
3. 80% $H_2O$ + 20% $H_2$: 4.92 NL·min$^{-1}$ $H_2O$ and 1.22 NL·min$^{-1}$ $H_2$
4. 90% $H_2O$ + 10% $H_2$: 5.5 NL·min$^{-1}$ $H_2O$ and 0.75 NL·min$^{-1}$ $H_2$
5. 45% $CO_2$ + 45% $H_2O$ + 10% $H_2$: 3.4 NL·min$^{-1}$ $CO_2$ + 3.4 NL·min$^{-1}$ $H_2O$ and 0.9 NL·min$^{-1}$ $H_2$
6. 25% $CO_2$ + 65% $H_2O$ + 10% $H_2$: 1.9 NL·min$^{-1}$ $CO_2$ + 4.92 NL·min$^{-1}$ $H_2O$ and 0.9 NL·min$^{-1}$ $H_2$

12 NL·min$^{-1}$ of oxygen or air was supplied to the oxygen electrode compartments. The use of a large variety of gas compositions was intended at identifying the response of individual electrodes. For co-electrolysis, the gas compositions were adjusted to yield a syngas mixture with a ratio suitable for production of methane, based on the thermodynamic equilibrium.

During the dynamic test, a constant flow of 65% $H_2O$ + 25% $CO_2$ + 10% $H_2$ corresponding to 23% $CO_2$ + 2% CO + 8% $H_2$ + 67% $H_2O$ in thermodynamic equilibrium composition was supplied to the fuel electrode compartment and pure oxygen was supplied to the oxygen electrode compartment. The gas composition to the fuel electrode was chosen aiming at the C–H ratio need for syngas production for $CH_4$ synthesis [31]. As the gas flow was kept constant during the cycling, the gas utilization changed, following the current density profile.

For dynamic operation, a simulated dynamic profile based on a daily wind profile as presented in Fig. 2 was used, and the test was done by controlling the supplied current to the stack. This profile was obtained from the island of Bornholm in Denmark from a 24 h period [5]. Naturally, there will be a fluctuation of the profile, changing over a whole year. However, the presently used profile is considered representative for the possible changes in magnitude and time. It was translated into an executable testing profile in order to imitate the storage requirements in the region. The daily wind profile was repeated for 1000 h testing. The maximum current density was set at 0.5 A cm$^{-2}$ in order to avoid high degradation. This current density limit was determined in durability tests of single cells [32].

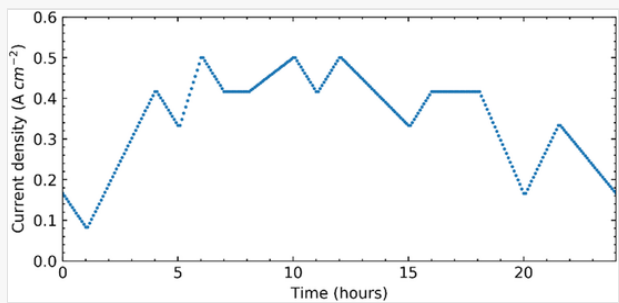

Fig. 2

Simulated daily wind profile for the stack test.

EIS measurements were carried out at both open circuit voltage (OCV) and under current (during dynamic testing) using a Solartron 1252A frequency analyzer with a Kepco BOP 50-4 M, which was used to boost the 16 mA AC current from the Solartron to 1A AC and together with an external current transducer in series with the cell. The spectra were recorded from 96850 to 0.08 Hz with 12 points per decade and were corrected using the short circuit impedance response of the test setup [33]. Due to the fact that the stack design was not optimized for impedance characterization, high frequency noise, which is caused by the stack geometry or impedance current/voltage cables, was observed in all the measured impedance spectra. In order to evaluate the stack resistance, high frequency (>25 kHz) data points were discarded and the ohmic resistance (serial resistance, Rs) was taken as the value of the real part of the impedance at 25 kHz. This is an alternative method to determine Rs [12]. The polarization resistance (Rp) was then calculated as the difference in the real part of the impedance at 25 kHz and 1 Hz. Analysis of the impedance data was performed using the software Ravdav [34].

## 3 Results

### 3.1 Initial performance-iV and EIS under co-electrolysis

The 6-cell stack was characterized for its initial performance at 750 °C prior to the durability testing. In Fig. 3, the i-V curves (Fig. 3a) and EIS (Fig. 3b) are presented for the 6 cells characterized under co-electrolysis conditions with $O_2$ supplied to the oxygen electrode compartment. The average cell open circuit voltage (OCV) was 0.893 V, close to the theoretical OCV 0.897 V calculated by the Nernst equation indicating a leak tight setup. All the cell voltages increased linearly with the increasing current density. The iV curves reveal a certain variation of the ASR among the six cells. The EIS results show very well agreement with the iV characterization, where Cell1 shows the largest ASR and Cell5 and Cell6 have the lowest ASR. However, all the 6 cells showed very similar gas concentration resistances (the low frequency arc in the right part of the EIS), which indicated an equal gas distribution inside the stack. On the other hand, EIS reveal that the variation of ASR among the cells is both due to a variation of the ohmic and the polarization resistances. A part of this variation might be due to temperature gradients or contact issues between the cells.

alt-text: Fig. 3

Fig. 3

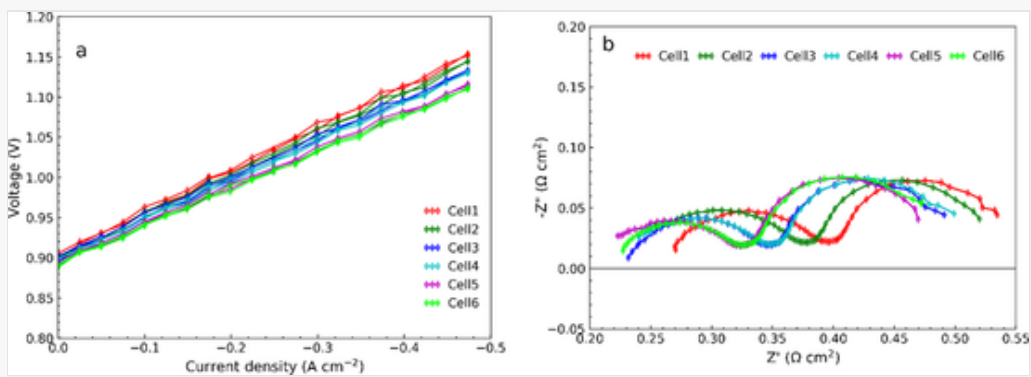

(a) i-V curves and (b) EIS for the 6 cells in the stack recorded with 65% $H_2O$ + 25% $CO_2$ + 10% $H_2$ to the fuel electrode and $O_2$ to the oxygen electrode at 750 °C prior to the durability test.

### 3.2 Durability test

Following the initial fingerprint, the stack was tested under dynamic load conditions. The profile for the current density along with cell voltage evolution is shown in Fig. 4a. Fig. 4b displays a zoom into 50 h for better visibility of the voltage evolution of the individual cells.

alt-text: Fig. 4

Fig. 4

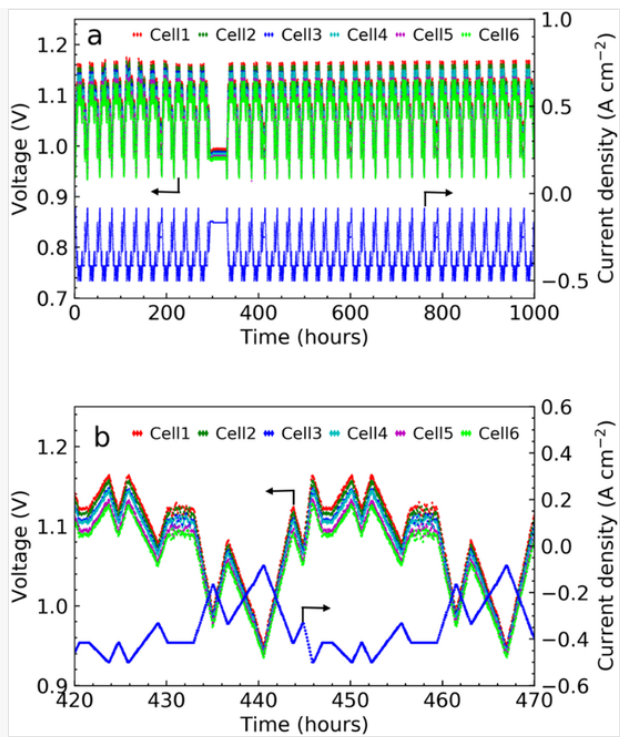

(a) Voltage evolution as a function of time during durability testing with 65% $H_2O$ + 25% $CO_2$ + 10% $H_2$ to the fuel electrode and $O_2$ to the oxygen electrode and current density (blue curve, right axis) modulation at 750 °C, (b) Voltage evolution for a selected period of 50 h of operation during durability testing with dynamic load. (For interpretation of the references to colour in this figure legend, the reader is referred to the Web version of this article.)

As can be seen from Fig. 4a and (b), the maximum cell voltage was approximately 1.2 V at a current density of 0.5 A cm$^{-2}$. Throughout the 1000 h of durability testing, no significant change in the stack voltage at the same current density was observed. The stack voltage exhibits an increase of only around 0.8%/1000 h.

During the dynamic load test, EIS were measured at 0.4 A/cm$^2$, i.e. always at the same current density, and are plotted in Fig. 5. Under this condition, the $H_2O$ + $CO_2$ utilization corresponds to 22%. The EIS only changed slightly for all the cells, illustrating that no significant degradation has occurred. Thus, the results confirm the findings from the voltage evolution over time (see Fig. 4).

alt-text: Fig. 5

Fig. 5

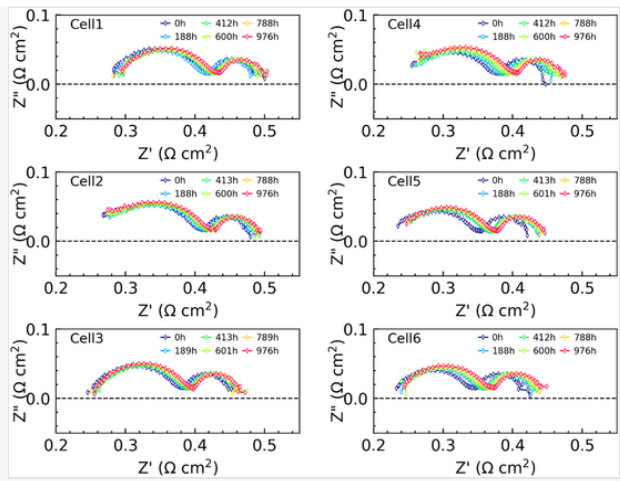

EIS measured on the single cells in the stack during the dynamic testing at −0.4 A cm$^{-2}$.

### 3.3 Detailed resistance evaluation

To evaluate the performance and degradation in detail, Rs, Rp and ASR values were extracted from the EIS measurements at −0.4 A/cm$^2$ during the durability test and are plotted in Fig. 6. Furthermore fuel inlet and outlet temperature are also plotted in Fig. 6a. During the first 200 h of operation under dynamic conditions, the resistances increased slightly. At the same time, the measured temperatures decreased. Therefore, the main part of the increase of resistances is most probably related to a temperature effect. Overall no significant change of ASR is visible and the linear change of ASR for all the cells is below 1% per 1000 h. No significant change of Rs is visible while Rp shows a very small linear increase over time during the durability test.

alt-text: Fig. 6

Fig. 6

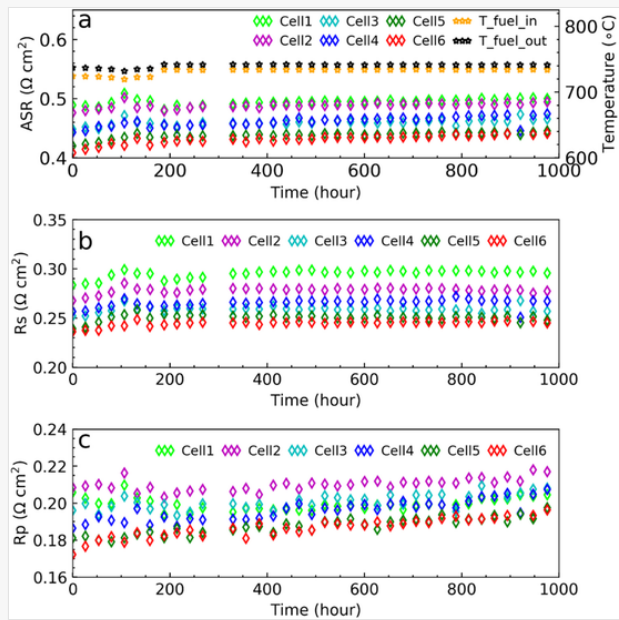

ASR (left axis), fuel inlet temperature and fuel outlet temperature (right axis) (a), $R_s$ (b) and $R_p$ (c) evolution during durability test, determined from EIS at 0.4 A/cm$^2$, with 65% $H_2O$ + 25% $CO_2$ + 10% $H_2$ applied to the fuel electrode and $O_2$ on the oxygen electrode at 750 °C.

To further investigate the cell degradation, the analysis of distribution of relaxation times (DRT) method was applied. In order to better break down the impedance contributions and associate them to the two electrodes, gas shift DRT analysis was performed using the gas shift impedance spectra measured under OCV prior to the dynamic load testing. OCV conditions provide typically a better resolution of the single electrode process contributions to the total impedance. By keeping the fuel electrode gas composition constant, the oxygen electrode gas was switched from air to oxygen. The response, in turn, is then related to the oxygen electrode. Similarly, by keeping the gas to the oxygen electrode constant and by varying the fuel electrode gas composition, the response of fuel electrode is indicated [35]. Fig. 7 a and b present the gas shift DRT analysis results of the fuel electrode and oxygen electrode, respectively, performed on cell number 3 of the stack, as a representative cell. Three fuel electrode process responses can be seen from Fig. 7a, with summit frequencies around 1 kHz, 40 Hz and 3 Hz, which may be attributed to the fuel electrode charge transfer process, gas diffusion process and gas conversion process, respectively. In the oxygen electrode gas shift DRT, one major process with a summit frequency around 100 Hz was found (Fig. 7b), representing oxygen electrode contributions. The DRT analysis results of the impedance measured at −0.4 A cm$^{-2}$ during the dynamic test are presented in Fig. 7c. It can be seen that for all the 6 measured cells, the major change of DRT took place at the frequency range around 1 kHz, which was identified from the gas shift analysis to be related to the fuel electrode charge transfer process. No significant oxygen electrode degradation process at a frequency around 100 Hz can be seen from Fig. 7c. Therefore, it can be concluded that the polarization resistance changes seen in Fig. 6c are mainly due to fuel electrode degradation.

alt-text: Fig. 7

**Fig. 7**

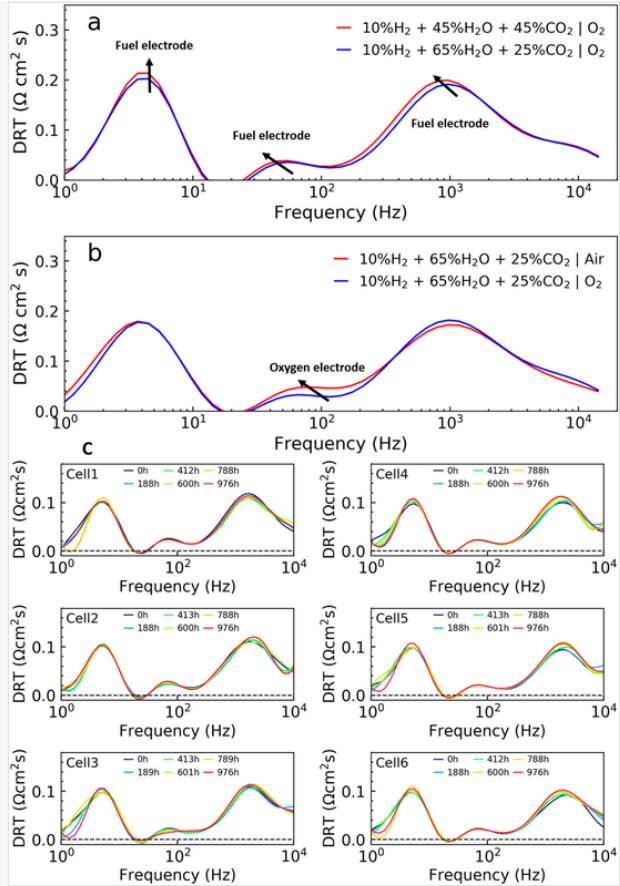

DRT plots showing frequency response (a) change of gas to fuel electrode, and (b) change of gas to oxygen electrode at OCV of Cell-3 of the stack (Arrows indicate the frequency ranges for the responses from both electrodes), (c) DRT plots of the cells in the stack recorded after selected times at 0.4 Acm$^{-2}$..

### 3.4 Syngas($H_2$+CO) production

During the co-SOEC process using steam and $CO_2$, synthesis gas (syngas: $H_2$ + CO) is formed. The production rate during the dynamic electrolysis mode can be calculated using the Faraday's law:

$$n = \frac{I}{2 \times F}$$

(1)

where $n$ is the gas production rate in mole $^{-1}$s, $I$ is the current in A and F is the Faraday's constant in C mol$^{-1}$. The stack power and syngas production rate as function of time during the dynamic test are plotted in Fig. 8. The power follows the current such that no significant change is observed thereby indicating nominal performance during 1000 h of operation. The accumulated syngas production throughout the 1000 h calculated based on the current and thermodynamic gas composition was approximately 3074 mol moles (258 m³ @ 25 °C).Thermodynamically, the ratio between $H_2$/CO varies with the changing of current density. Nevertheless, a theoretical ratio of $H_2$/CO = 3.7 can be achieved based on the calculation from thermodynamic equilibrium and the average applied current.

### Fig. 8

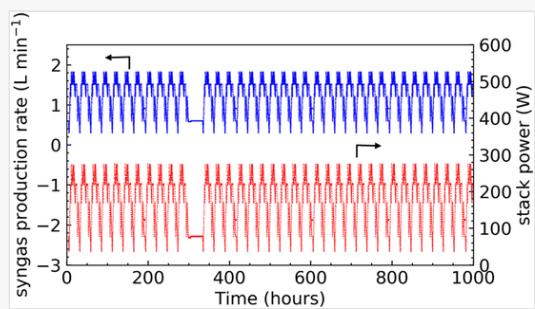

Syngas production rate (left axis) and stack power (right axis) as function of time during durability testing.

## 4 Discussion

From the initial i-V curves as shown in Fig. 3a, a certain variation of performance of the cells is observed, which is also visible from EIS during initial characterization as displayed in Fig. 3b.

For a better comparison, ASR obtained from the i-V curves and EIS are listed in Table 1. The values from the i-V curves are taken in the linear regime as the slope of the curve from 0 to 0.25 A cm$^{-2}$. Cell-1 has a higher Rs than the other cells, leading to a lower performance in terms of ASR. This may be related to the assembly of the cells in the stack leading to a potentially additional contact resistance for Cell1 or temperature effects in the stack. All the other five cells have similar Rs, but there is a certain variation of the polarization resistances as well. Cell1 has the highest ASR, which agrees with the EIS measurements. Cell1 has a higher initial voltage, which also remains higher than for all the other cells throughout the testing as seen in Fig. 4. However, the overall variation of performance among the cells is very small. The ASR obtained from the EIS agrees well with the ASR calculated from the iV curves, which gives a good validation of the two different characterization methods, particularly the more challenging EIS analysis of the cells within the stack environment.

### Table 1

> ⓘ The presentation of Tables and the formatting of text in the online proof do not match the final output, though the data is the same. To preview the actual presentation, view the Proof.

ASR values during initial characterization with 65% $H_2O$ + 25% $CO_2$ + 10% $H_2$ applied to the fuel electrode and $O_2$ applied to the oxygen electrode at 750 °C.

| Cell number | ASR from i-V curves ($\Omega$ cm$^2$) | ASR from EIS ($\Omega$ cm$^2$) |
|---|---|---|
| 1 | 0.54 | 0.55 |
| 2 | 0.53 | 0.54 |
| 3 | 0.50 | 0.50 |
| 4 | 0.51 | 0.51 |
| 5 | 0.49 | 0.49 |
| 6 | 0.49 | 0.50 |

During the dynamic test, except the first 200 h, where the Rs change is expected to be due to the decreasing of the gas inlet temperature, no significant change of Rs is observed, indicating a good contact within the stack, despite the dynamic operating conditions, and the absence of degradation of the ohmic resistance. A slight increase of Rp is observed over time, especially on Cell4, Cell5, and Cell6. DRT analysis of the EIS recorded at the same current density of $-0.4$ A cm$^{-2}$ as presented in Fig. 7c shows that the major change of the resistance took place in the high frequency region, with a summit frequency of around 1 kHz. According to the gas shift DRT analysis (Fig. 7a and b), this frequency region is attributed to the fuel electrode process. Minor change in the middle frequency around 50–200 Hz is only seen on Cell2 and Cell3, however, due to the frequency overlapping between the oxygen electrode process and fuel gas diffusion process, it is hard to solely attribute the degradation to one of the two possible processes. It has to be underlined that those changes are very small.

To comprehend the contribution from electrodes to the total ASR, the overpotential was calculated according to the following equation [36]:

$$Op = V - OCV - iRs \tag{2}$$

Where, Op is the overpotential, V is the measured voltage under current, OCV is the measured OCV, i is the operating current density and Rs is the serial resistance. It is also possible to calculate the overpotential from the current and the polarization Rp extracted from the EIS measurement. However, such a method requires that the changing of the Rp is linear with current, otherwise, integration of Rp as function of current has to be performed to calculate the overpotential, which requires more measurement points of Rp from 0 A to the operation current. Therefore in this manuscript Equation (2) is applied, since Rs is independent of current density (assuming no temperature change), thereby ohmic law can be applied.

The overpotential for all the cells of the stack at a current density of 0.4 A cm$^{-2}$ was calculated at the beginning and the end of the durability test and is displayed in Table 2. Beginning of the test refers to 257 h, since only the responses at similar temperature can be directly compared.

alt-text: Table 2

Table 2

The presentation of Tables and the formatting of text in the online proof do not match the final output, though the data is the same. To preview the actual presentation, view the Proof.

Overpotential of the cell (excluding serial resistance).

| Cell number | Overpotential (mV)(Beginning of the durability test, t = 257 h) | Overpotential (mV)(End of the durability test, t = 1000h) | Change of overpotential (%) |
| --- | --- | --- | --- |
| 1 | 82.2 | 86.3 | 5 |
| 2 | 97 | 100.9 | 4 |
| 3 | 104.6 | 108.4 | 3.7 |
| 4 | 100.4 | 109.5 | 9 |
| 5 | 111.7 | 113.4 | 1 |
| 6 | 117.6 | 120 | 2 |

The overpotentials of the 6 cells in the stack differ with around 35 mV at the same current density of 0.4 A cm$^{-2}$. Cell1 has the highest ASR and exhibits the lowest overpotential, while Cell6 with lowest ASR exhibits the highest overpotential. This may be due to the fact that the major difference of Cell1 and Cell6 lies in their polarization resistance, which agrees well with the Rp plot as shown in Fig. 6c. Nevertheless, the differences of the initial overpotentials did not result in a large difference in the degradation or final overpotential, and, in fact, there were only small increases of overpotential values after the durability test. Previous studies have reported the loss of active Ni network due to Ni migration as a major contribution to the degradation of such cells [18]. However, the maximum current density in this study was limited to 0.5 A cm$^{-2}$ in order to prevent such processes. This means that the overpotential experienced by the cells is in the window of no or reversible degradation, where no major Ni particle rearrangements or losses should occur [16]. By running the stack under dynamic load, the cells in the stack were exposed to various current densities but not to high current density (high overpotential) for a long time, which led to small degradation. Therefore, the magnitude of applied current density seems to be the determining factor for degradation and not the dynamic mode of changing of the current density. This provides the opportunity to identify optimal operating windows based on durability tests at constant operating conditions and at single cell level. While keeping cell/stack operation within this safe window, the long term conversion of excess renewable energy into syngas was demonstrated.

## 5 Conclusion

In this work, a SOEC stack was operated in co-SOEC mode under dynamic conditions, where the current density was modulated following a simplified wind profile simulating electricity input from fluctuating sources. In the applied current density profile, the maximum was 0.5 A cm$^{-2}$ at 750 °C. Under these conditions, cell & stack degradation was insignificant. ASR degradation of less than 1%/1000 h was observed. Analyzing EIS during stack operation at one selected current density of 0.4 A cm$^{-2}$ revealed a constant gas conversion impedance for all the cells of the stack, indicating even distribution of the gas flows. The very small change of the frequency response at around 1 kHz with only minor degradation effects, most probably related to the fuel electrode TPB reactions process. The formed synthesis gas can be further converted into for example methane through well-known catalytic processes and stored and distributed in the existing infrastructure as synthetic natural gas. The results thus demonstrate at stack level the capabilities of SOEC to utilize electricity from fluctuating sources for syngas production, which can be further converted to storage media or used for fuels in the transport sector.

## Declaration of competing interest

The authors declare that they have no known competing financial interests or personal relationships that could have appeared to influence the work reported in this paper.

## Acknowledgements

The authors wish to thank Mr. H. Henriksen, Mr. Ole Hansen, Dr. J. Høgh, Dr. S. Pitscheider, Dr. N. Seselj and Dr. A. Ploner for their technical help. The research leading to these results has received funding from the European Union's Horizon 2020 framework program (H2020) for the Fuel Cells and Hydrogen Joint Technology Initiative under grant agreement no. 699892 ("Efficient Co-Electrolyser for Efficient Renewable Energy Storage-ECo").

## Glossary

| | |
|---|---|
| ASR | Area specific resistance, a measure of resistance to the performance |
| DRT | Distribution of relaxation times, analysis method for degradation |
| Durability | Long-term performance |
| EIS | Electrochemical impedance spectroscopy, analysis tool |
| LSC-GDC | Lanthanum strontium cobaltite-gadolinium doped ceria, SoA oxygen electrode material for SOECs |
| Ni-YSZ | Nickel-yttria stabilized zirconia, SoA fuel electrode material for SOECs |
| NL/min | Normal liter per minute, standard flow measurement unit |
| OCV | Open circuit voltage, measured cell voltage under zero net current |
| Overpotential | Change in electrode potential from when the electrode is in equilibrium |
| PtG/PtL | Power to-Gas/Power-to-Liquid, technologies aimed at electricity conversion to chemicals |
| SoA | State-of-the-art, current technology in use |
| SOEC | Solid oxide electrolysis cell, technology for converting steam and $CO_2$ to useful chemicals |
| Syngas | a mixture of carbon monoxide and hydrogen ($CO+H_2$), a valuable precursor for methane |

# Highlights

- Short stack operated for the first time in co-electrolysis mode under dynamic conditions.
- ASR degradation rates lower than 1%/1000 h for over 1000 h operation.
- Main degradation associated to fuel electrode processes.